# Enhanced refrigerant capacity and magnetic entropy flattening using a two-amorphous FeZrB(Cu) composite

Pablo Álvarez[1], José L. Sánchez Llamazares[2], Pedro Gorria[1,a)], and Jesús A. Blanco[1]

[1]Departamento de Física, Universidad de Oviedo, Calvo Sotelo, s/n, 33007 Oviedo, Spain

[2]División de Materiales Avanzados, IPCyT, Camino a la presa San José 2055, CP 78216, San Luis Potosí, Mexico

The temperature dependence of the isothermal magnetic entropy change, $\Delta S_M$, and the magnetic field dependence of the refrigerant capacity, $RC$, have been investigated in a composite system $x$A+(1-$x$)B, based on $Fe_{87}Zr_6B_6Cu_1$ (A) and $Fe_{90}Zr_8B_2$ (B) amorphous ribbons. Under a magnetic field change of 2 T the maximum improvement of the full-width at half maximum of $\Delta S_M(T)$ curve (47 and 29 %) and the $RC$ (18 and 23 %), in comparison with those of the individual alloys (A and B), is observed for $x \approx 0.5$. Moreover, a flattening over 80 K in the $\Delta S_M(T)$ curve around room temperature range is observed, which is a key feature for an Ericsson magnetic refrigeration cycle.

75.30.Sg, 75.50.Bb, 75.50.Kj, 75.40.Mg

a)Electronic mail: pgorria@uniovi.es

Nowadays the search for new materials with enhanced magneto-caloric effect (MCE) for their utilization in room temperature magnetic refrigerators is a very active field of research due to its lower energy consumption and environmental friendly character [1]. The ideal Ericsson cycle (two isothermal and two isomagnetic field processes) is optimal for RT applications [2]. Its maximum efficiency is reached when the MCE exhibits a constant temperature dependence of the isothermal magnetic entropy change, $\Delta S_M(T)$, within the operating temperature range [3]. This condition is difficult to be accomplished by a single material, but a composite made of two ferromagnetic materials may fulfill it provided that the difference between their Curie points, $T_C$, is customized [4]. However, sintered mixtures of compounds are not well suited because they behave as a material with a single $T_C$ [5]. Instead of that, the design of specific composites [6-8] can lead to an enlargement of the MCE temperature span and an almost constant $\Delta S_M(T)$ curve. The efficiency of the magnetic material in terms of the energy transfer between the cold ($T_{cold}$) and hot ($T_{hot}$) reservoirs, is quantified by its refrigerant capacity, RC, an important figure of merit that characterizes the magneto-caloric material [9,10]:

$$RC(H) = \int_{T_{cold}(H)}^{T_{hot}(H)} \left[\Delta S_M(T)\right]_H dT \qquad (1)$$

where $T_{cold}$ and $T_{hot}$ are commonly selected as the temperatures corresponding to the full width at half maximum of $\Delta S_M(T)$. In order to attain large RC values, a large magnetic entropy change over a wide temperature range is desirable. However, it has been shown that although the maximum value of $\Delta S_M$ decreases, a large broadening of the $\Delta S_M(T)$ curves can overcompensate such decrease, giving rise to a significant *RC* enhancement [7,10]. In turn, numerical calculations and their proof-of-principle in practical systems show that the design and practical realization of multiphase or composite magneto-caloric materials with optimized $RC$ and $\Delta S_M(T)$ curves is not a simple task [4-6]. For a two-phase composite the magnetic field variation of the $RC$ is complex, and depends on several factors such as $\Delta T_C$ and the shape, width and peak value of $\Delta S_M(T)$ curve of the two constituents [6].

In this letter we report on the MCE in a two-ribbon composite showing a larger $RC$ with respect to that of each individual ribbon and a flattened $\Delta S_M(T)$ curve. The composite is formed by two Fe-rich FeZrB(Cu) amorphous ribbons. These alloys, which were extensively studied due to their re-entrant spin-glass [11] and invar behaviors [12], are particularly appropriate since, depending on the Fe content, the $T_C$ is tunable between 200 and 400 K [13,14]. FeZrB(Cu) alloys exhibit broad $\Delta S_M(T)$ curves with moderate peak values, $\Delta S_M^{peak}$, ($\approx 1.5$ J kg$^{-1}$ K$^{-1}$ for a magnetic field change of $\mu_o \Delta H = 2.0$ T) [6,15,16]; thus offering the possibility of mixing two alloys with the appropriate $\Delta T_C$ difference.

Amorphous ribbons with nominal composition $Fe_{87}Zr_6B_6Cu_1$ (A) and $Fe_{90}Zr_8B_2$ (B) were produced by melt-spinning. The composite consists of two ribbons with approximate dimensions of 1 cm (long) × 1 mm (wide) × 25 μm (thick) glued together with a Kapton® film. Magnetization measurements were performed using a vibrating sample magnetometer (Quantum Design). A set of $M(H)$ curves was measured from 100 to 400 K each 10 K up to $\mu_o H = 5.0$ T for each material, with the magnetic field applied parallel to the ribbon largest length (1 cm) in order to minimize the demagnetizing factor. The temperature dependence of $\Delta S_M$ was calculated by numerical integration of the adequate Maxwell relation [9]. The total magnetic entropy change for the two-ribbon composite, $\Delta S_{comp}(T)$, can be calculated from the individual $\Delta S(T)$ curves [17]:

$$\Delta S_{comp}(T,H,x) = x\Delta S_A(T,H) + (1-x)\Delta S_B(T,H) \qquad (2)$$

where $x$ and $(1-x)$ are the weight amount of ribbons A and B, respectively in the composite system $x$A+$(1-x)$B.

For $\mu_o\Delta H$ = 2.0 T, $x \approx 0.5$ maximizes the $RC$ value and $\Delta S_{comp}(T)$ exhibits a flattened region or table-like behavior. For these reasons, a combination with $x = 0.5$ was selected to compare the $\Delta S_{comp}(T)$ curves with those obtained from A and/or B ribbons [$\Delta S_A(T)$, $\Delta S_B(T)$ or $0.5\Delta S_A(T)+ 0.5\Delta S_B(T)$ curves].

Figs. 1 (a) and (b) show the low field $M(T)$ and the $\Delta S_M(T)$ curves (for $\mu_o\Delta H$ = 2.0 T and 5.0 T) for A and B ribbons. The Curie temperatures are $T_{C,A}$ = 300(5) K and $T_{C,B}$ = 240(5) K, hence, $\Delta T_C$ = 60 K [13]. The $\Delta S_{comp}(T)$ curves show a broad single-peak for all $x$ [see Fig. 1(c)], however, for low values of $\mu_o\Delta H$ the $\Delta S_{comp}(T)$ curve exhibits a double peak shape [see inset of Fig. 2(a)] and $RC_{comp}$ cannot be defined. Moreover, the central region of $\Delta S_M(T)$ becomes relatively flat and the maximum of $\delta T_{FWHM}$ (full width at half maximum of the $\Delta S_M(T)$ curve) occurs for $x \approx 0.5$ [see inset of Fig. 2(b)].

For layered composite materials, the value of $RC_{comp}$ and the attainment of a constant $\Delta S_M$ inside a certain temperature interval depend strongly on the applied magnetic field [4,6]. Fig. 2(a) depicts how $RC_{comp}$ varies with $x$ in our composite system for $\mu_o\Delta H$ = 2.0 and 5.0 T. The maximum value of $RC_{comp}$ corresponds to $x \approx 0.5$, and is larger than those $RC$ values for either A or B (see upper right panel in fig. 3). This $RC$ enhancement is a consequence of the increase in $\delta T_{FWHM}$. Nevertheless, a compromise between the value of the $\delta T_{FWHM}$ and the potential lost of efficiency of the machine (due to an increase of cycles in the heat exchange medium) is needed. In turns, the $\Delta S_{comp}(T)$ curves under low magnetic field changes ($\mu_o\Delta H$ < 0.4 T) exhibit a double-peak profile [see inset of Fig. 2(a)], and the $\delta T_{FWHM}$ cannot be properly defined.

Both, the enhancement of $RC_{comp}$ and the flattening observed in the $\Delta S_{comp}$ curve stimulated us to measure the $M(H)$ curves for the two-ribbon composite. This is important in order to assess whether the shape of $M(H)$ curves is in some extend affected by dipolar interactions between the individual components A and B. In Fig. 3 the $\Delta S_{comp}(T)$ curves obtained from the $M(H)$ curves measured on the composite through numerical calculation of the Maxwell relation [3,9], and by eq. (2) using $\Delta S_M(T)$ of the individual constituents are compared. The excellent agreement between both curves suggest that the dipolar interaction between the ribbons can be neglected, and validates the $\Delta S_{comp}(T)$ curves shown in Fig. 1(c).

On the other hand, the flattening, or temperature range where the $\Delta S_M$ remains almost constant, is an important attribute of the system if an ideal Ericsson refrigerant cycle is considered, since it avoids the generation of irreversible work [3]. Assuming a 10% difference with respect to $|\Delta S_{comp}^{peak}|$, the $\Delta S_{comp}(T)$ curve flattens in the intervals 220-300 K, and 230-320 K, for 2.0 and 5.0 T respectively. Finally, the broadening of the $\Delta S_{comp}(T)$, due to the increase of the $\delta T_{FWHM}$ in more than 30 K (see left upper panel in Fig. 3), gives rise to an improvement of the $RC$ (see Fig. 3 upper right panel).

We have shown that the full-width at half maximum of the $\Delta S_M(T)$ curve in FeZrB amorphous alloys may be significantly increased via a layered composite made of two ribbons with different values of $T_C$. As a consequence, a significant improvement of the refrigerant capacity $RC_{comp}$ can be obtained. For $\mu_o\Delta H = 2.0$ T, we found increments of 47 and 29 % (18 and 23 %) in $\delta T_{FWHM}$ ($RC_{comp}$) compared with the individual A and B ribbons, respectively. In addition, $\Delta S_{comp}(T)$ showed a flattening over 80 K around room temperature, which is a key attribute that magnetic coolants must possess for its potential use in an Ericsson cycle. These findings path the way to an enhancement of the room temperature MCE through the adequate combination of materials with second order magnetic phase transitions. Therefore, we foresee that these results will attract the attention of researchers currently in charge of developing magnetic refrigeration devices that can exploit the present results.

**Acknowledgments**. We thank Spanish MICINN and FEDER program for financial support under MAT2008-06542-C04-03 research project. JLSL acknowledges the support received from Conacyt, Mexico, under the project number 156932.

FIGURE CAPTIONS

Figure 1. (color online) (a) Magnetization vs. temperature curves under $\mu_oH$ = 5 mT. (b) $\Delta S_M(T)$ curves of samples A and B for $\mu_oH$ = 2 and 5 T. (c) $\Delta S_{comp}(T)$ curves for the two-ribbons system calculated from the isothermal $M(H)$ curves measured independently for samples A and B for $\mu_oH$ = 2 and 5 T.

Figure 2. (color online) (a) $RC_{comp}$ as a function of $x$ for $\mu_oH$ = 2 and 5 T. Inset: illustration of the difficulty encountered to define $RC$ for a double-peak $\Delta S_{comp}(T)$ curve. (b) $\delta T_{FWHM}$ vs. $x$ for $\mu_oH$ = 2 and 5 T.

Figure 3. (color online) $\Delta S_{comp}(T)$ curves for the composite 0.5A+0.5B ($\mu_oH$ = 2 and 5 T) obtained from the measured $M(H)$ curves of the composite (full symbols) and from eq. (2) (open symbols). Left upper panel: Experimental magnetic field dependence of the $\delta T_{FWHM}$ for the samples A, B and the composite system. Right upper panel: $RC$ vs. $\mu_oH$ for the samples A, B and the composite.

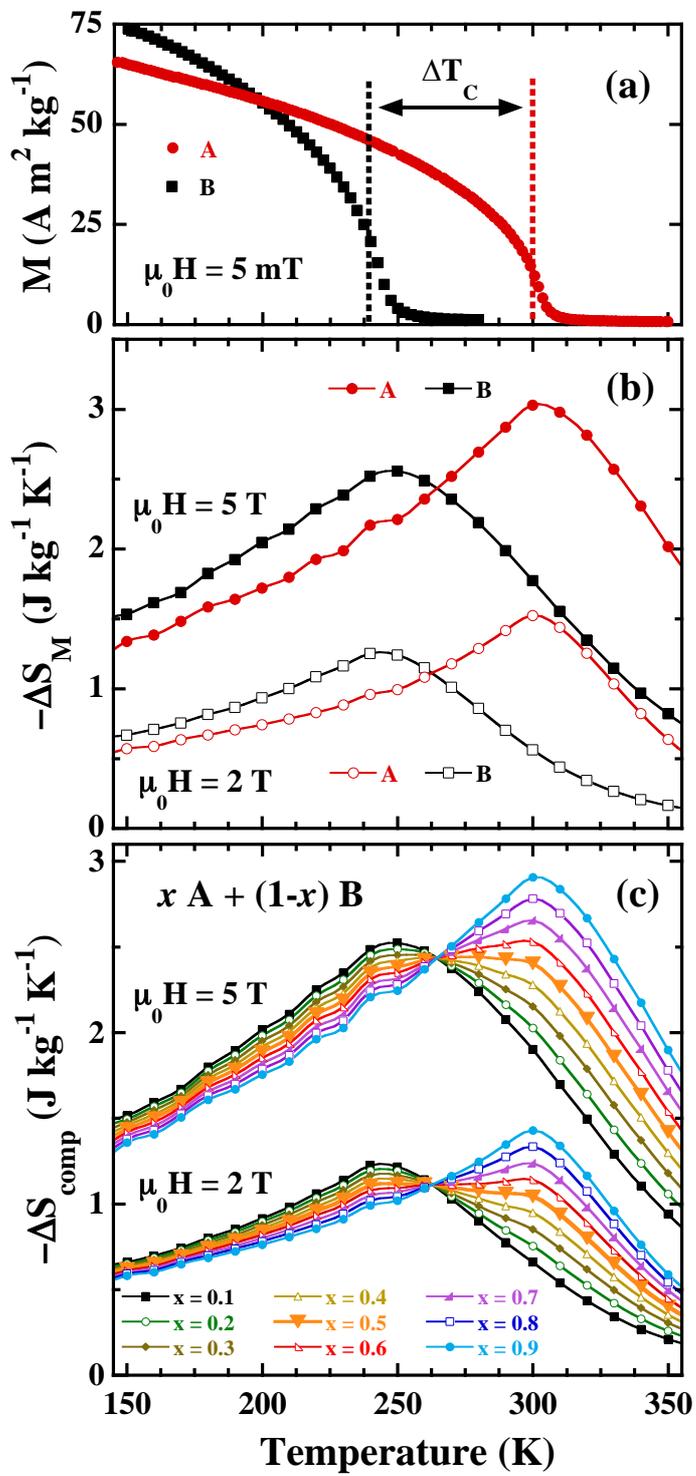

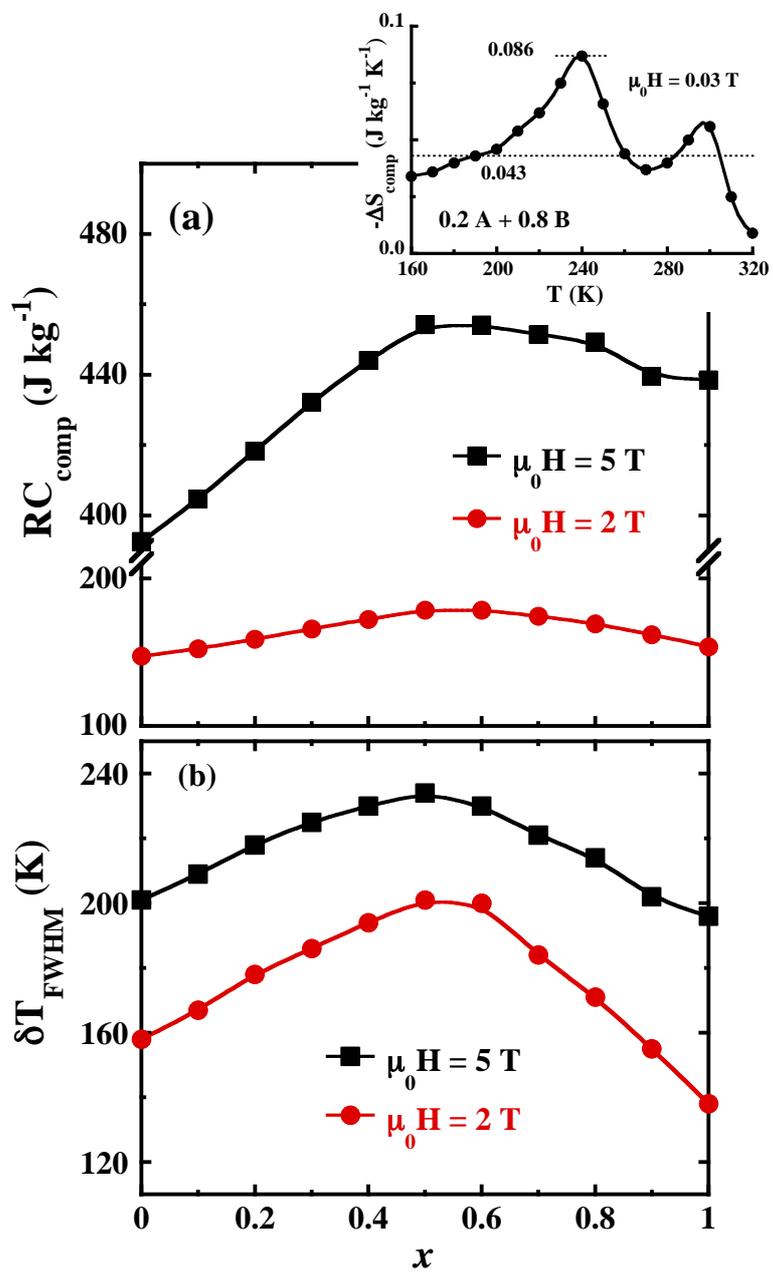

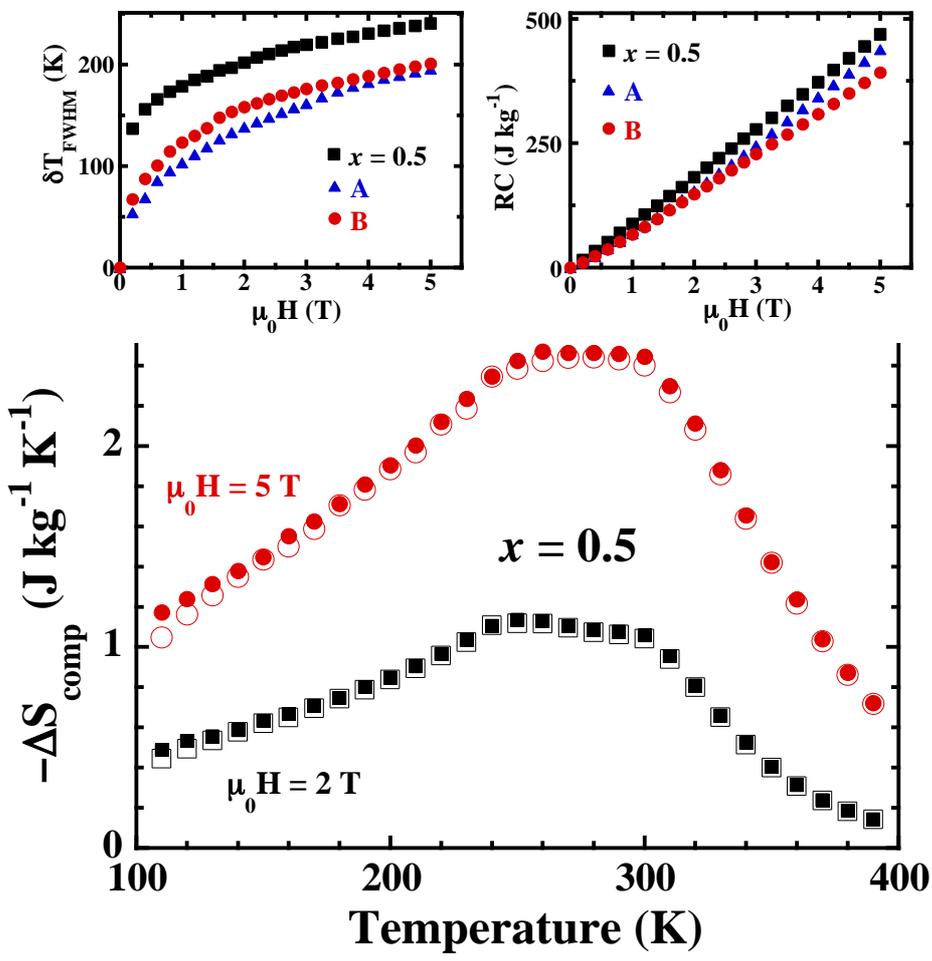